\documentclass[letterpaper, 10 pt, conference]{ieeeconf}

\IEEEoverridecommandlockouts
\usepackage{graphicx}
\usepackage{siunitx}
\usepackage{caption}
\usepackage{subcaption}
\usepackage{placeins}
\usepackage{dblfloatfix}

\title{\LARGE \bf A New Wind Farm Active Power Control Strategy to Boost Tracking Margins in High-demand Scenarios}

\author{Simone Tamaro$^{1}$ and Carlo L. Bottasso$^{1}$%
\thanks{$^{1}$Technical University of Munich, Wind Energy Institute, Boltzmannstrasse 15, 85748, Garching, Germany
        {\tt\small simone.tamaro@tum.de}, {\tt\small carlo.bottasso@tum.de}
        \newline \newline 2023 American Control Conference (ACC), San Diego, CA, USA, 2023, pp. 192-197, doi: 10.23919/ACC55779.2023.10156275.\newline© 2023 IEEE.  Personal use of this material is permitted.  Permission from IEEE must be obtained for all other uses, in any current or future media, including reprinting/republishing this material for advertising or promotional purposes, creating new collective works, for resale or redistribution to servers or lists, or reuse of any copyrighted component of this work in other works.}%
}
\begin{document}

\maketitle
\thispagestyle{empty}
\pagestyle{empty}

\begin{abstract}

This paper presents a new active power control algorithm designed to maximize the power reserve of the individual turbines in a farm, in order to improve the tracking accuracy of a power reference signal. The control architecture is based on an open-loop optimal set-point scheduler combined with a feedback corrector, which actively regulate power by both wake steering and induction control. The methodology is compared with a state-of-the-art PI-based controller by means of high-fidelity LES simulations. The new wind farm controller reduces the occurrence of local saturation events, thereby improving the overall tracking accuracy, and limits fatigue loading in conditions of relatively high-power demand.

\end{abstract}

\section{INTRODUCTION}

The  growth of wind energy penetration in the electricity mix requires new control algorithms to keep the electrical grid in balance~\cite{NREL,Bridging_The_Gaps}. When operating in active power control (APC) mode, a wind farm  intentionally extracts less than the available power from the wind, in order to meet the demands of the transmission system operator (TSO). The application of APC to a wind farm is not trivial and introduces new challenges. In fact, the maximum available power dependents on ambient conditions, which vary dynamically in uncertain ways~\cite{vkuik16}. Additionally, wind may suddenly drop, possibly leaving not enough power reserves to track a given reference signal~\cite{fleming2016}. In a wind farm, the situation is further complicated by the presence of low-momentum turbulent wakes, which are responsible for power losses and fatigue loading of waked turbines~\cite{VERMEER2003467,lee2012}. Various solutions have been proposed to mitigate wake effects, such as induction and yaw control~\cite{JMeyers_etAl_2022}. The latter consists of ``steering'' the wake  away from downstream rotors, and its effectiveness for power boosting has been demonstrated numerically~\cite{ballasto}, experimentally in the wind tunnel~\cite{Campagnolo_2016}, as well as in field trials~\cite{fleming19,bart21}.

Different APC approaches have been presented in the literature. An open-loop APC strategy is discussed in~\cite{fleming2016}. The authors showed that the lack of feedback poses a limitation on the power tracking accuracy of the method, especially in conditions of strong waking. Furthermore, an equal dispatch of power sharing among the turbines proved to be suboptimal, due to the different local power reserves induced by the heterogeneity of the flow.

Recently, various authors have used model predictive control (MPC) for APC~\cite{Shapiro17,Boersma_2018,vali18a}. The main drawback of such methods lies with the need of a dynamic farm flow model, which can be computationally expensive.

Simpler control structures based on classical PI (proportional integral) loops have also been extensively investigated~\cite{VANWINGERDEN20174484}. While lacking the sophistication of MPC, such methods do not need a wind farm flow model and can provide fast response times with simple implementations. The APC PI controller of ref.~\cite{VANWINGERDEN20174484} operates on the tracking error and adjusts the power demands to follow a reference, sharing power in an arbitrary, static manner among the turbines. The method includes gain scheduling based on the fraction of saturated wind turbines, defined as the ones whose available power is smaller than the demanded one. This method was improved in ref.~\cite{vali19} by dynamically adjusting the set-points of the wind turbines, with the goal of equalizing their loading. The authors tested this methodology with an actuator disk model using large eddy simulations (LES). Later, this approach was also demonstrated with the more sophisticated actuator line method (ALM) in LES~\cite{silva21}. So far these  PI-based methods have been applied only to induction control, and they are not necessarily optimal. Moreover, saturation conditions are problematic, due to the possible local lack of power reserves (margins), which are not explicitly accounted for nor monitored in the existing implementations.

In this paper, a new wind farm control architecture is presented to improve the power tracking accuracy in conditions of strong persistent wakes, when the wind farm power demand is close to the maximum available power. An improved tracking performance is obtained by explicitly maximizing the power margin, in order to hedge against wind lulls. This novel methodology combines wake steering with induction control. Wake steering is used because of its ability to increase power margins by mitigating wake effects~\cite{Campagnolo_2016}. Wake steering is implemented through an open-loop model-based set-point optimal scheduler, closely following the standard implementation that has recently become popular in power-boosting wind farm control~\cite{JMeyers_etAl_2022}. Induction control is implemented through a fast closed-loop corrector to improve tracking accuracy. The new methodology is demonstrated  in a partial wake impingement scenario of a cluster of turbines, using a TUM-modified version of NREL's ALM-LES Simulator fOr Wind Farm Applications (SOWFA)~\cite{FLEMING2014211,JW}.%,CW1}.

The paper is structured as follows. First, the novel APC methodology is presented. Second, the simulation model is described and finally, results are discussed for steady-state and unsteady conditions.

\begin{figure*}[b!]
  \centering
%  \framebox{\parbox{3in}{\includegraphics[scale=0.2]{Figures/sketchers.png}}}
  \includegraphics[width=0.75\textwidth]{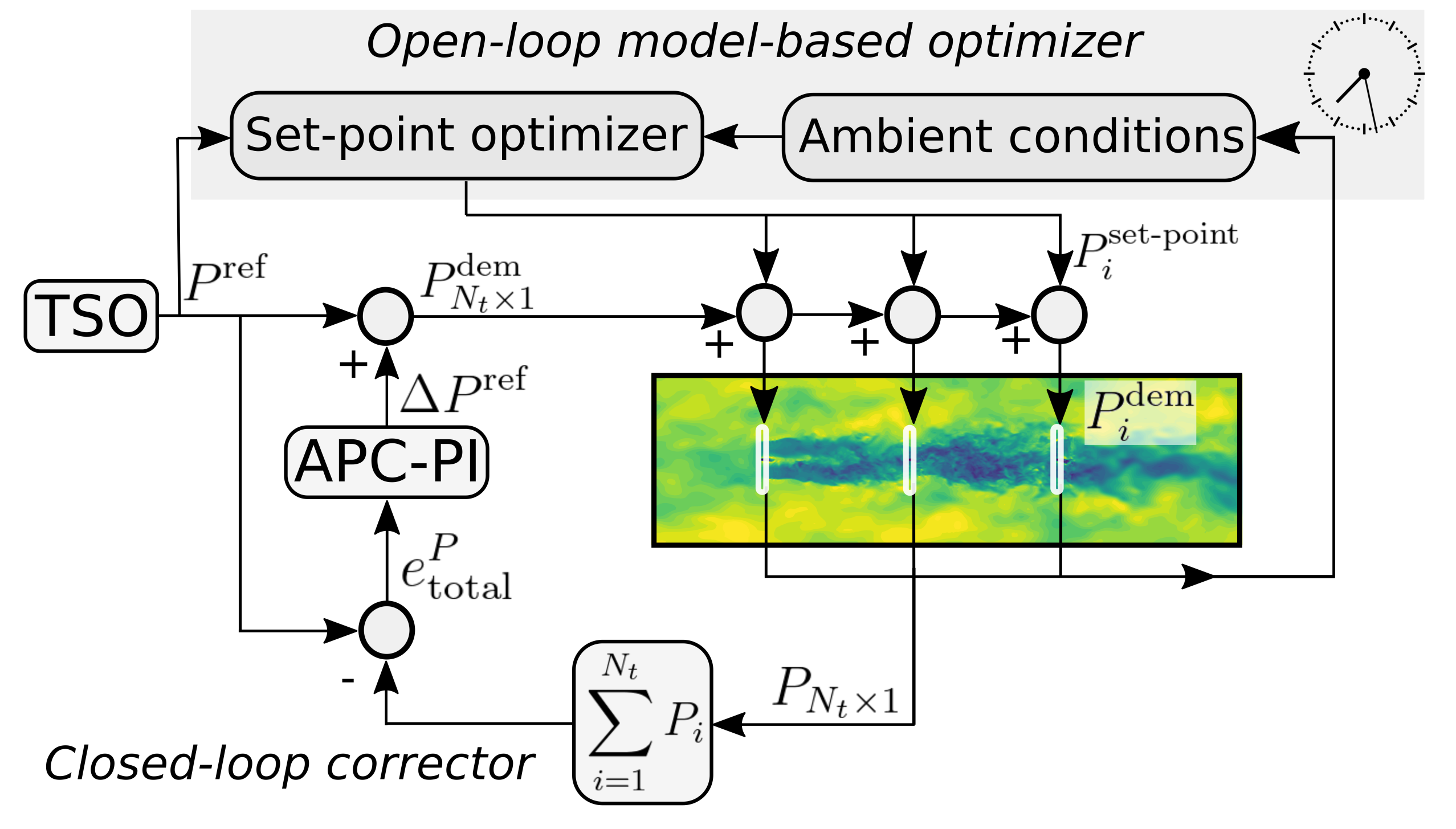}
\caption{Schematic representation of the APC controller, featuring an open-loop model-based optimizer and a closed-loop corrector.}
\label{sketch1}
\end{figure*}

\section{METHODOLOGY}\label{meth}

The core of the proposed wind farm control architecture is an open-loop model-based set-point optimal scheduler. This control element determines the yaw misalignment of each turbine and its contribution to the demanded value (i.e. power share), given the power demand required by the TSO and the ambient conditions. The latter can be obtained in real time from SCADA data or with wind sensing methods~\cite{marta21}. A feedback loop serves the main purpose of correcting tracking errors, which will inevitably arise from the open loop control element during operation. A sketch of the overall control architecture is shown in fig.~\ref{sketch1}. The closed and the open loops are executed at two distinct time rates, since their outputs involve physical phenomena characterized by different time scales. Specifically, the open loop updates the yaw-set points and the power shares at a slower rate, due to the time required by the wake to propagate downstream. On the other hand, the closed loop changes the turbine inductions at a faster pace, to reduce tracking errors.

\subsection{Open-loop set-point optimal scheduler}
The open-loop component of the algorithm provides the optimal set-points in terms of yaw misalignment and power share. These are computed by a gradient-based optimization that maximizes the smallest power reserve within the wind turbines of the farm, for a given overall power demand.

The power of the $i$th turbine is noted $P_i = P_i (A_i,u_i)$, where $A_i$ indicates the local ambient conditions (here assumed to include wind speed, wind direction and turbulence intensity), and $u_i$ are the control inputs (namely, induction and yaw misalignment). Power is computed using a wind farm flow model, which here is based on the FLOw Redirection and Induction in Steady-state (FLORIS v2) tool~\cite{gebraad16}.

The maximum power that can be captured by turbine $i$ by adjusting its control set-point $u_i$ (while keeping the set-points of the other turbines fixed) is computed as
\begin{displaymath}
P_{a,i} = \textrm{arg} \, \max_{u_i} P_i (A,u_i) = \frac{1}{2}\rho \pi R^2 C_p U^3 \cos^{P_p}(\gamma),
\end{displaymath}
where $\rho$ is the air density, $R$ is the wind turbine radius, $U$ is the undisturbed free-stream velocity, and $P_p$ is the cosine exponent relating the yaw misalignment angle $\gamma$ to power. The algorithm looks for the combination of set-points that produce the maximum possible minimum power ratio $P_i/P_{a,i}$ across all turbines in the farm, while satisfying the power demand of the TSO. This can be expressed as
\begin{equation}\label{fminimax}
\min_u \, \max_{i \in [1,N]} \frac{P_i}{P_{a,i}}\text{
such that }\sum_{i=1}^{N} P_i=P_\textrm{ref}.
\end{equation}
In fact, the smaller the power ratio $P_i/P_{a,i}$, the larger the margin ${m_i = 1-P_i/P_{a,i}}$ that is available to compensate against drops in the wind. Equation~(\ref{fminimax}) represents a constrained optimization problem, which is solved with the gradient-based Sequential Quadratic Programming (SQP) method~\cite{sqp79}. The optimization does not need to be performed in real time during operation. Rather, it is executed offline for a set of ambient conditions and relative wind farm capacities. Results are collected in a look-up table, which is then interpolated at run-time, similarly to what is routinely done for power-boosting wind farm control~\cite{JMeyers_etAl_2022}.

In the example shown later in this work, the open loop is executed every 30~seconds.

\subsection{Closed-loop corrector}
The closed-loop corrector is directly taken from the work of ref.~\cite{VANWINGERDEN20174484}, and it is executed every 0.01~seconds. The corrector consists of a simple PI feedback loop that operates on the power tracking error, which arises from the open-loop component of the control structure. The tuned PI gains used in this work are $K_{P,APC}=0.2$ and $K_{I,APC}=\SI{0.05}{\second^{-1}}$.

\subsection{Identification of saturation conditions}
On each turbine, the occurrence of saturations is determined by a condition that combines tracking error and  pitch angle. In particular, a saturation is detected when the blade pitch is at its optimal value and the tracking error exceeds a given negative threshold, set to the value of \SI{100}{\kW} in this work. The magnitude of this threshold determines the aggressiveness of the wind farm controller. This method was chosen because it can be implemented based on standard information that is readily available on board wind turbines, and does not rely on uncertain and difficult-to-estimate parameters such as thrust coefficient or axial induction.

\section{NUMERICAL MODEL}
\subsection{Steady-state model}
The engineering farm flow model FLORIS v2~\cite{gebraad16} is used here both to synthesize the open-loop part of the controller and to perform steady-state analyses, prior to testing in the dynamic higher-fidelity LES-ALM environment. The standard FLORIS implementation is extended with the option to derate the turbines by modifying the $C_p$ and $C_t$ tables, following a basic curtailment approach. Moreover, a linear dependency of the power loss exponent $P_p$ with $C_t$ is also included in the model~\cite{cossu21,heck22}, so that
\begin{displaymath}
P_p=A\,C_t+B,
\end{displaymath}
where $A=-1.56$ and $B=3.16$, based on experimental and numerical observations. This dependency between the power loss exponent and the thrust coefficient is particularly relevant when combining derating and yaw misalignment, since the wind turbines operate at a wide range of $C_t$ values due to their dynamic curtailment.

\subsection{Unsteady simulations}
\begin{figure}[hbtp]
\begin{center}
\includegraphics[width=0.475\textwidth]{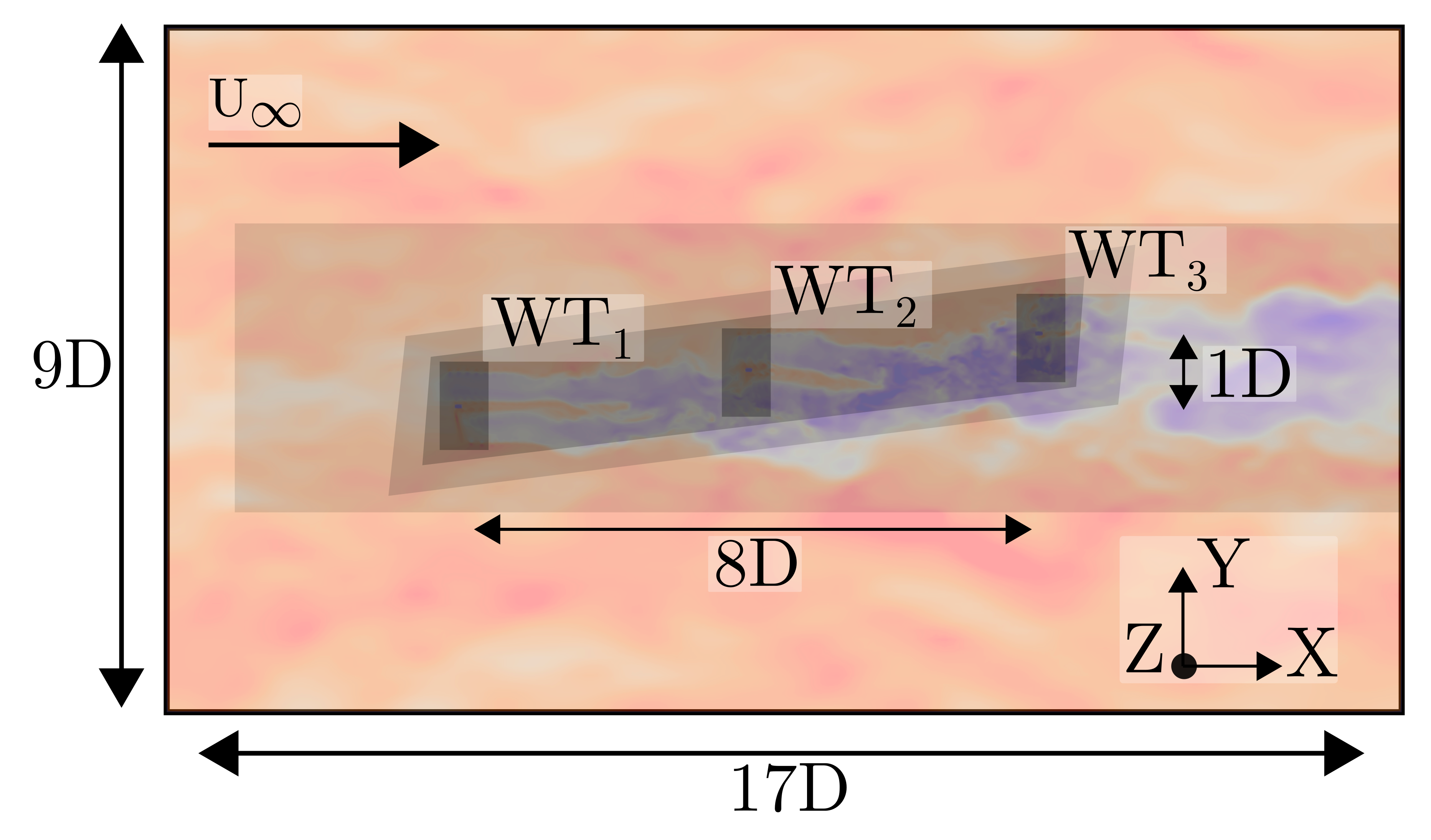}
\caption{Wind farm layout and simulation scenario. The shaded areas indicate the mesh refinement levels.}
\label{mesh}
\end{center}
\end{figure}

LES-ALM simulations are used for testing the performance of the new APC formulation, because they are able to deal with the complex dynamics typical of wind turbine wakes and their interactions~\cite{JW}. %,CW1}.
The filtered ALM of refs.~\cite{Troldborg_2007,meneveau_2019} is used to model the blades, by projecting forces computed along the lifting lines onto the LES mesh grid. Simulations are run with a turbulent wind obtained from a precursor generated in stable atmospheric conditions. The Cartesian mesh consists of approximately 13.5 million cells, and uses six refinement levels. The smallest cells measure \SI{1}{\meter}, and are located in correspondence of the rotors. The computational domain, grids and turbine layout are shown in fig.~\ref{mesh}.

\section{RESULTS AND ANALYSIS}
The scenario analyzed in this paper consists of a cluster of three IEA \SI{3.4}{\MW} wind turbines~\cite{iea335mw}, installed at a distance of 4 diameters and misaligned by half a diameter relatively to the incoming wind vector. The scenario is adapted from~\cite{Campagnolo_2016}, and it is chosen to mimic the typical operating conditions of an onshore wind plant with close spacings and partial wake overlaps. The inflow is characterised by a turbulence intensity of 6\% at hub height, a shear of 0.2, and a mean wind speed of \SI{9.5}{\meter\per\second}, equal to the rated speed of the turbines.

\subsection{Steady-state conditions}\label{steady_state}
\begin{figure}[h!]
     \centering
     \begin{subfigure}[b]{0.475\textwidth}
         \centering
        \includegraphics[width=0.75\textwidth]{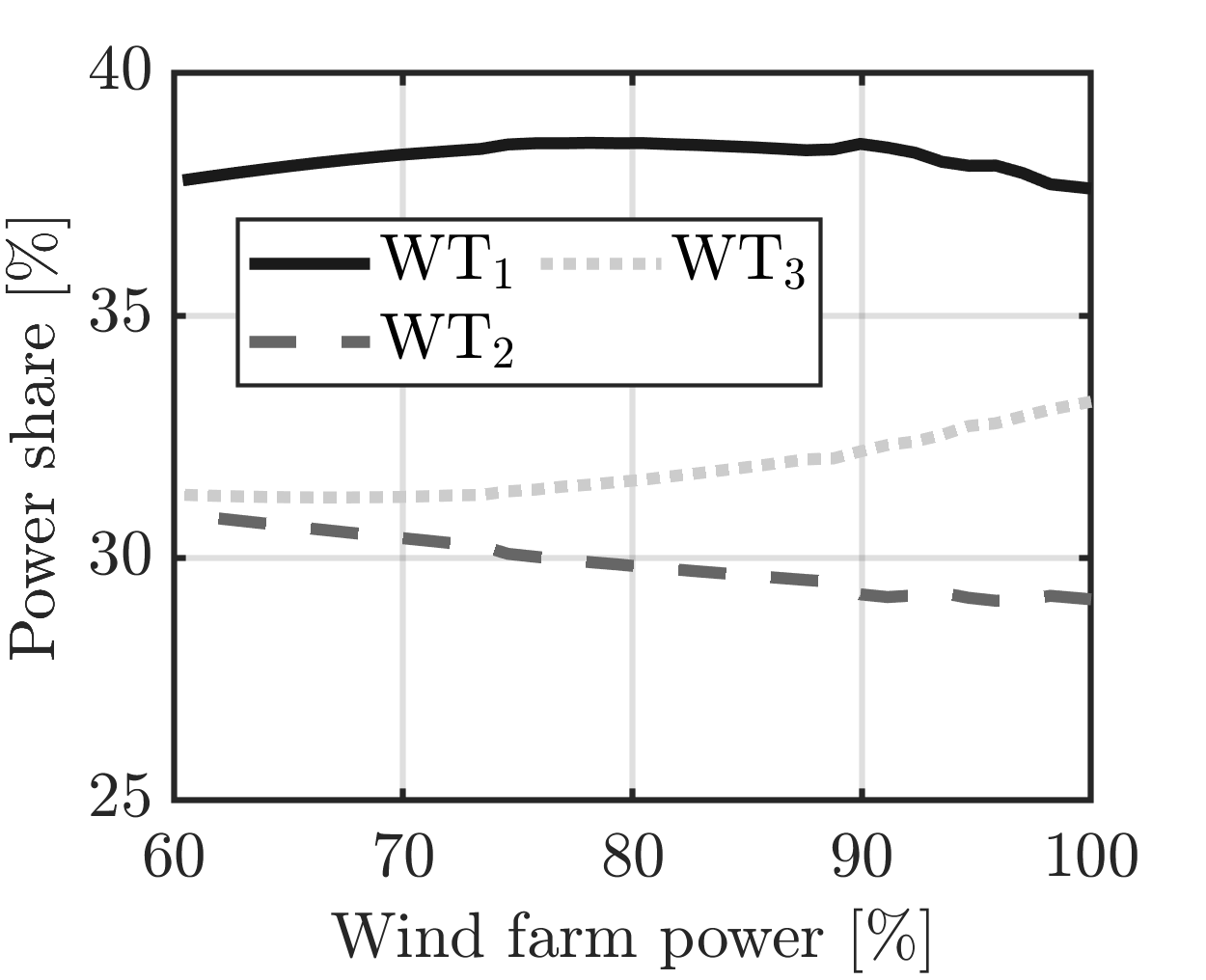}
         \caption{Power share as a function of wind farm power.}
         \label{fig4a}
      \end{subfigure}
      \par\bigskip
     \begin{subfigure}[b]{0.475\textwidth}
         \centering
        \includegraphics[width=0.75\textwidth]{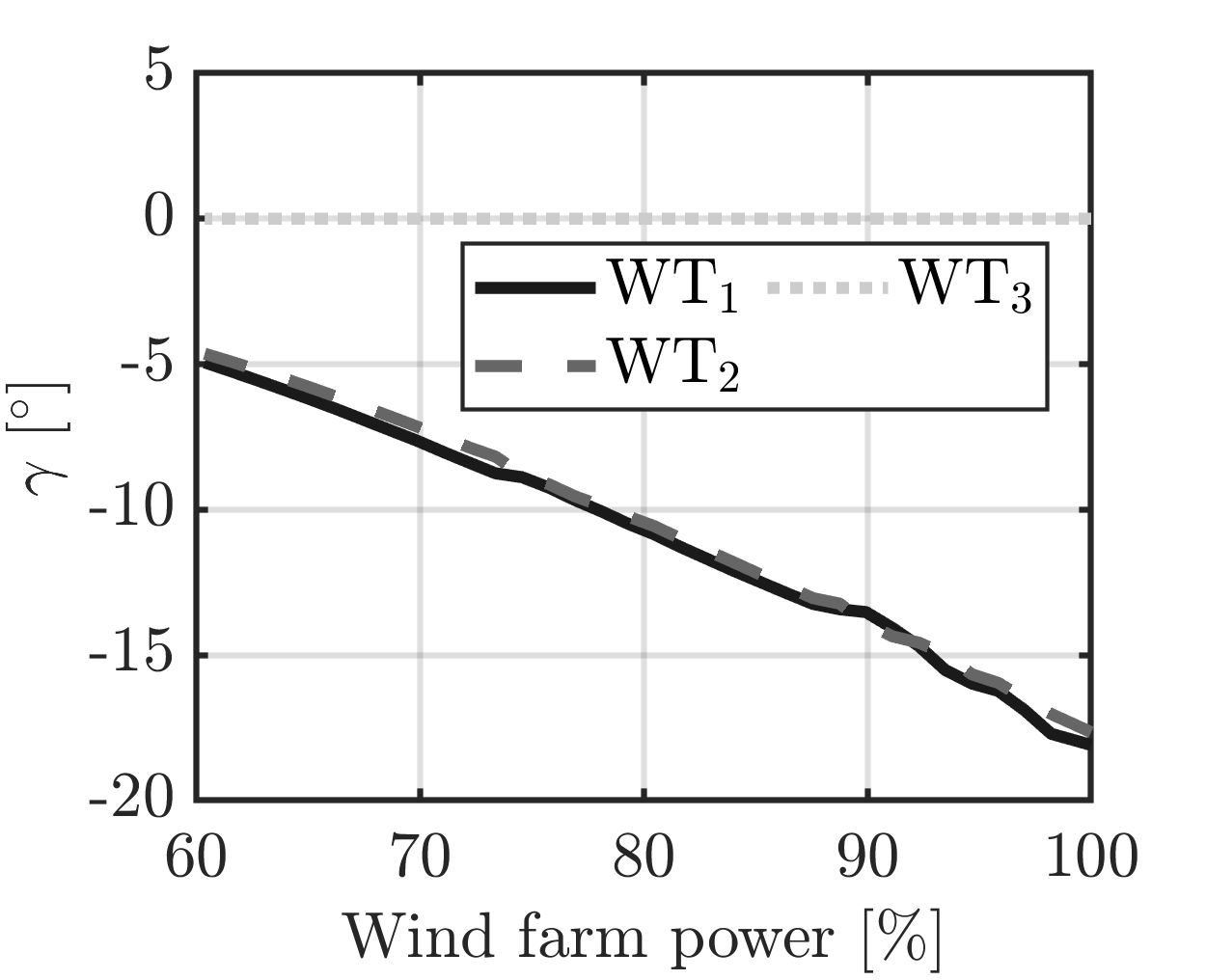}
        \caption{Yaw set-points as a function of wind farm power.}
         \label{fig4b}
     \end{subfigure}
\caption{Optimal set-points that maximize the minimum power margin.}
\label{fig4}
\end{figure}

First, the open-loop optimal scheduler is demonstrated in steady-state conditions. For each turbine, fig.~\ref{fig4} reports the yaw set-points and power share percentage that maximize the smallest power margin.

The figure shows that the most upstream turbines are misaligned relatively to the wind, with the goal of  increasing the power reserves of the downstream ones. Moreover, power share is not distributed equally, because of different local inflow conditions and wake effects.

\begin{figure}[htbp]
\begin{center}
\includegraphics[width=0.425\textwidth]{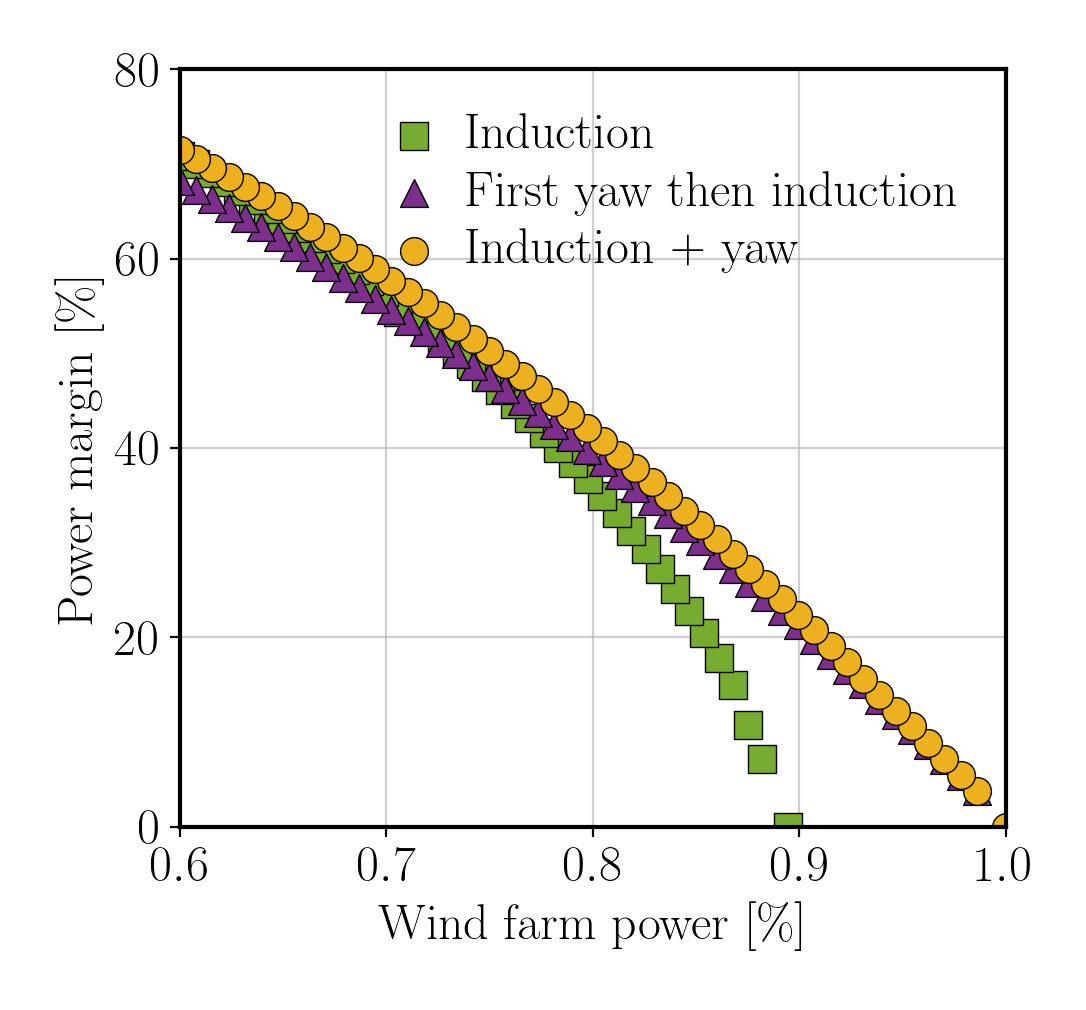}
\caption{Smallest local power reserve in percentage obtained with three wind farm control strategies.}
\label{margins}
\end{center}
\end{figure}

These margin-optimal set-points (noted \textit{induction + yaw} in the following) are compared to the ones of two alternative strategies in fig.~\ref{margins}. In the first of these strategies (noted \textit{induction}), only induction is used to match the demand (i.e. the turbines are always aligned with the incoming wind vector). In the second (noted \textit{first yaw then induction}), the turbines are first misaligned to maximize power capture, and then induction control is used to match the demand. In both cases, the power share is computed in order to maximize the smallest power margin in the wind farm.

The figure shows that ---as expected--- the margin drops to zero in correspondence of the maximum power of the plant, and increases as the power demand is lowered and the wind turbines are derated. Compared to the \textit{induction} case, the methods featuring wake steering are able to significantly increase the power margin for a wide range of wind farm power demands. Furthermore, the \textit{first yaw then induction} strategy generates similar margins to the \textit{induction + yaw} case at relatively high TSO demands. However, its performance drops slightly as the power demand is lowered, because of the power losses caused by its larger persistent yaw misalignments. These losses are particularly enhanced by the low thrust coefficient at which the turbines operate, due to curtailment~\cite{cossu21,heck22}. Because of its better ability to generate large margins, only the \textit{induction + yaw} strategy is considered in the remainder of this work.

\subsection{Unsteady simulations}
Next, the methodology is tested with unsteady CFD simulations. Results are compared with the controller developed in ref.~\cite{vali19}, which is assumed here as the state-of-the-art benchmark.

A dynamic reference power signal typical of automatic generation control (AGC) is used as input signal. AGC is the secondary response regime of grid frequency control, and it consists in the modification of the power output of a plant depending on the dynamically changing requests by the transmission system operator~\cite{NREL}. A similar signal has been considered by other authors~\cite{fleming2016,Shapiro17,Boersma_2018,VANWINGERDEN20174484,vali19}.

\begin{figure}[htbp!]
     \centering
     \begin{subfigure}[b]{0.475\textwidth}
         \centering
            \includegraphics[width=\textwidth]{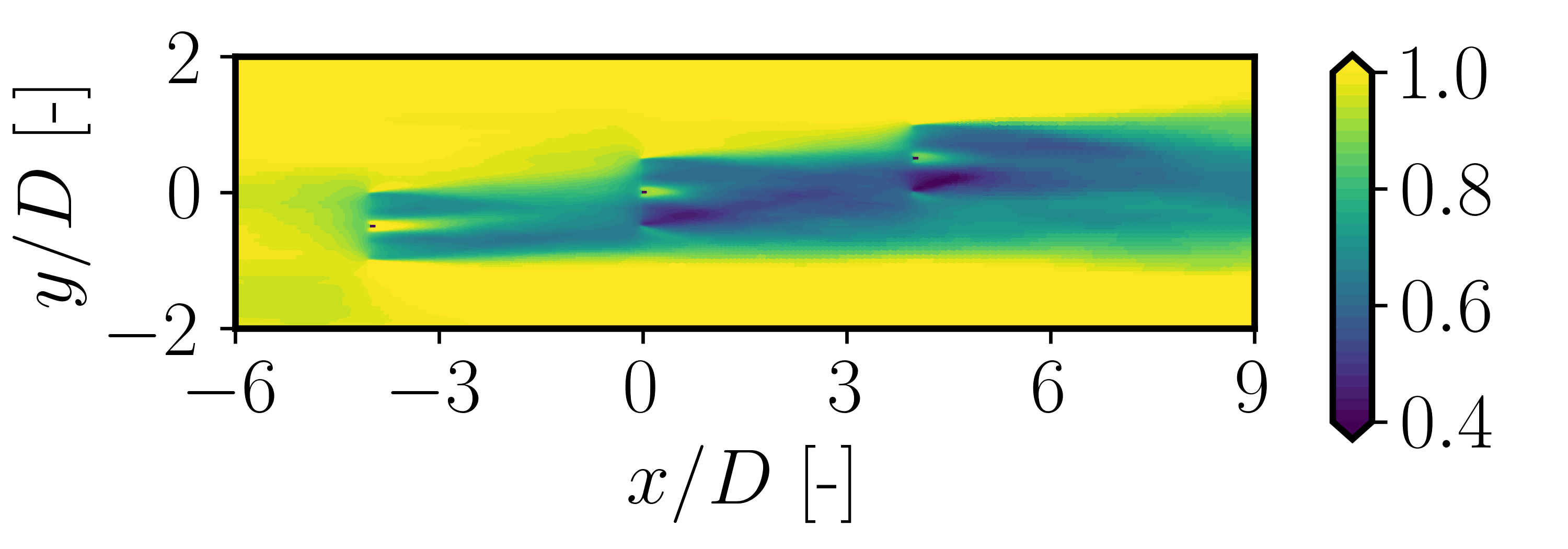}
        \caption{Vali et al., 2019}
        \label{}
      \end{subfigure}
     \begin{subfigure}[b]{0.475\textwidth}
         \centering
            \includegraphics[width=\textwidth]{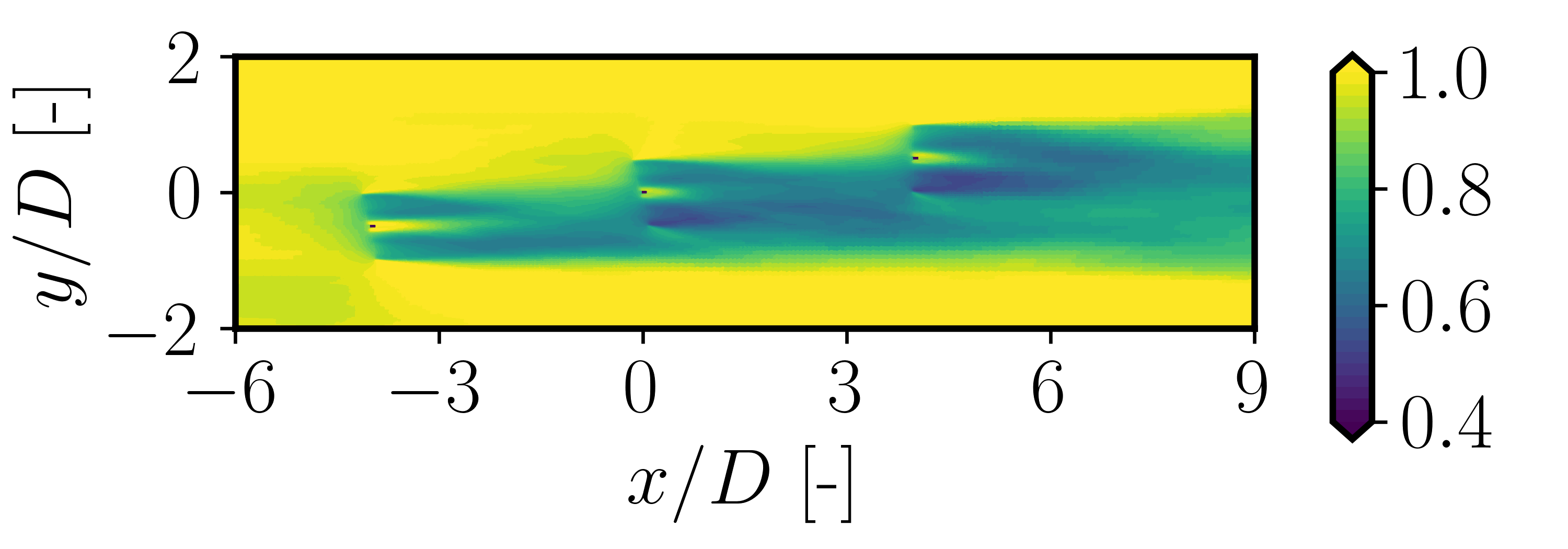}
            \caption{Induction + yaw}
         \label{averageFields_b}
     \end{subfigure}
 \caption{Mean streamwise velocity fields, non-dimensionalised by the free-stream wind speed.}
 \label{averageFields}
\end{figure}

Fig.~\ref{averageFields} presents the average velocity fields in the wind farm obtained with the benchmark control and with the proposed induction+yaw approach. The effect of yaw misalignment can be clearly observed, as the wakes of the upstream turbines appear to have been deflected in Fig.~\ref{averageFields_b}.

\begin{figure}[htbp]
\centering
\includegraphics[width=0.425\textwidth]{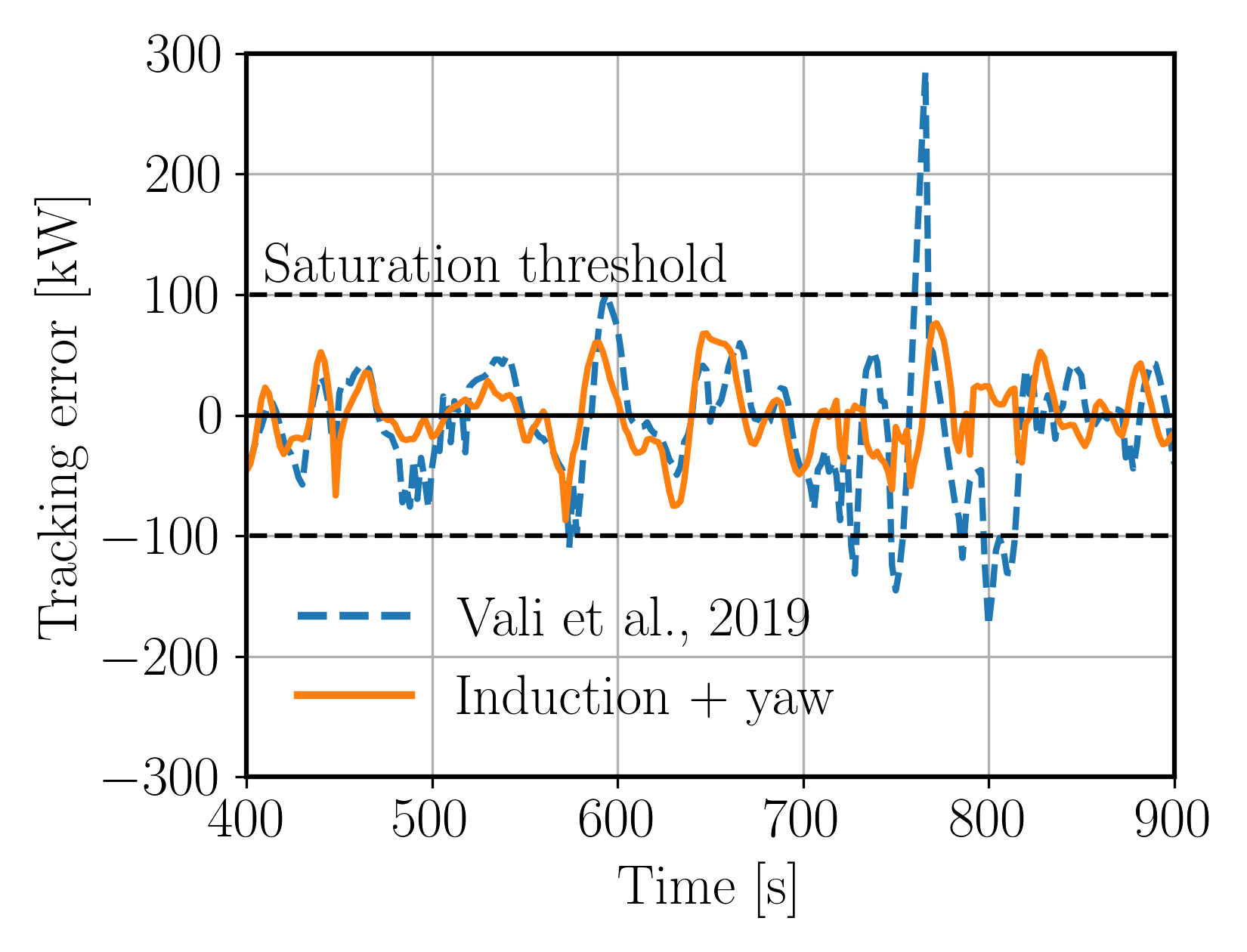}
\caption{Power tracking error vs. time, for the proposed (solid orange line) and benchmark (dashed blue line) controllers.}
\label{trackerror}
\end{figure}

Fig.~\ref{trackerror} shows a comparison of the power tracking error obtained with the benchmark method and the newly proposed one.

The figure shows that the benchmark method presents frequent negative deviations from the reference signal. These deviations are due to the power saturation of the wind turbines operating in waked inflow conditions. On the other hand, the controller featuring wake steering is capable of reducing the frequency of occurrence of these phenomena, thereby improving the overall tracking accuracy. For the results of fig.~\ref{trackerror}, the new wind farm controller reduces the root-mean-square of the tracking error by 42.6\% relatively to the benchmark. In the latter, the significant error occurring at $t\approx\SI{760}{\second}$ is due to a simultaneous saturation of all the wind turbines in the cluster.

\begin{figure}[htbp!]
     \centering
     \begin{subfigure}[b]{0.425\textwidth}
         \centering
            \includegraphics[width=\textwidth]{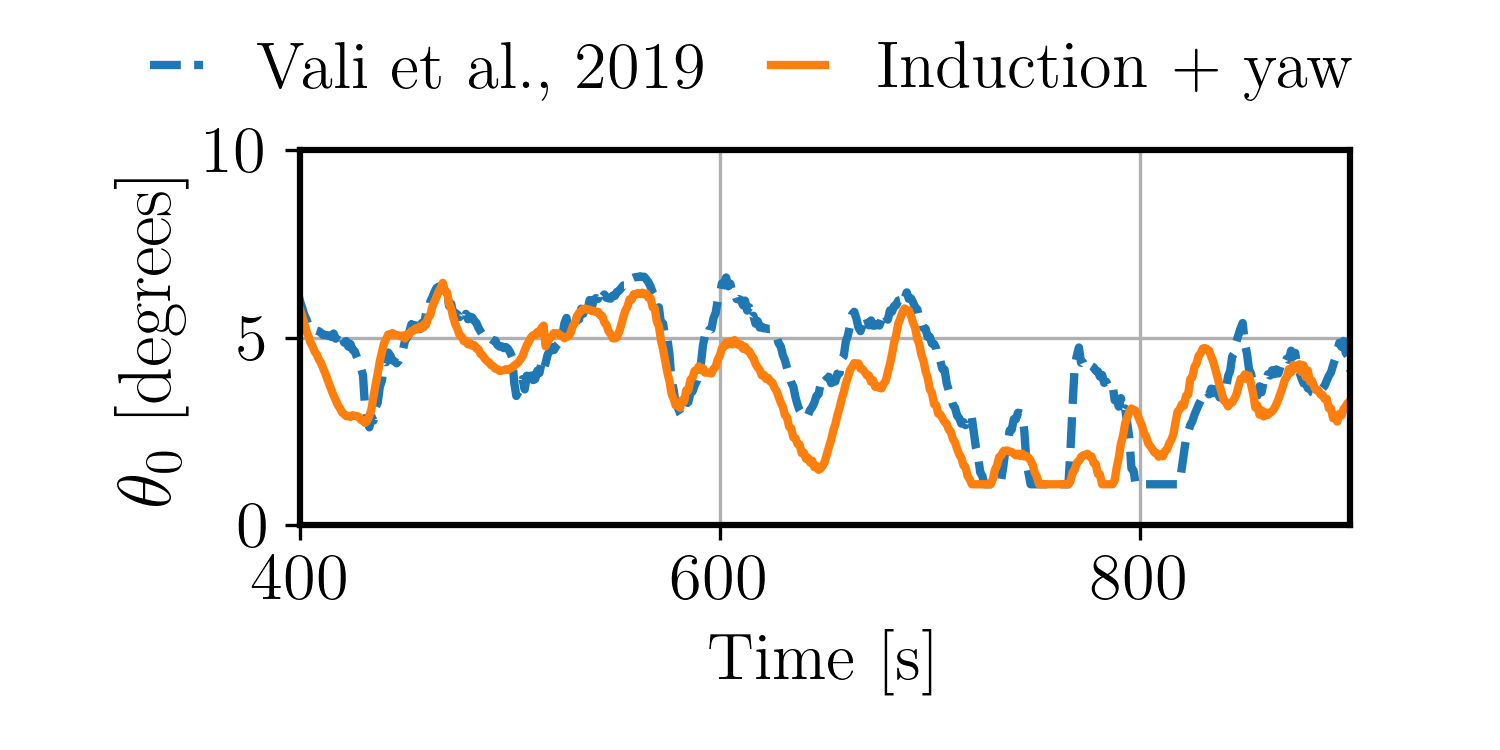}
        \caption{Wind turbine 1 (upstream)}
        \label{6a}
      \end{subfigure}

     \begin{subfigure}[b]{0.425\textwidth}
         \centering
            \includegraphics[width=\textwidth]{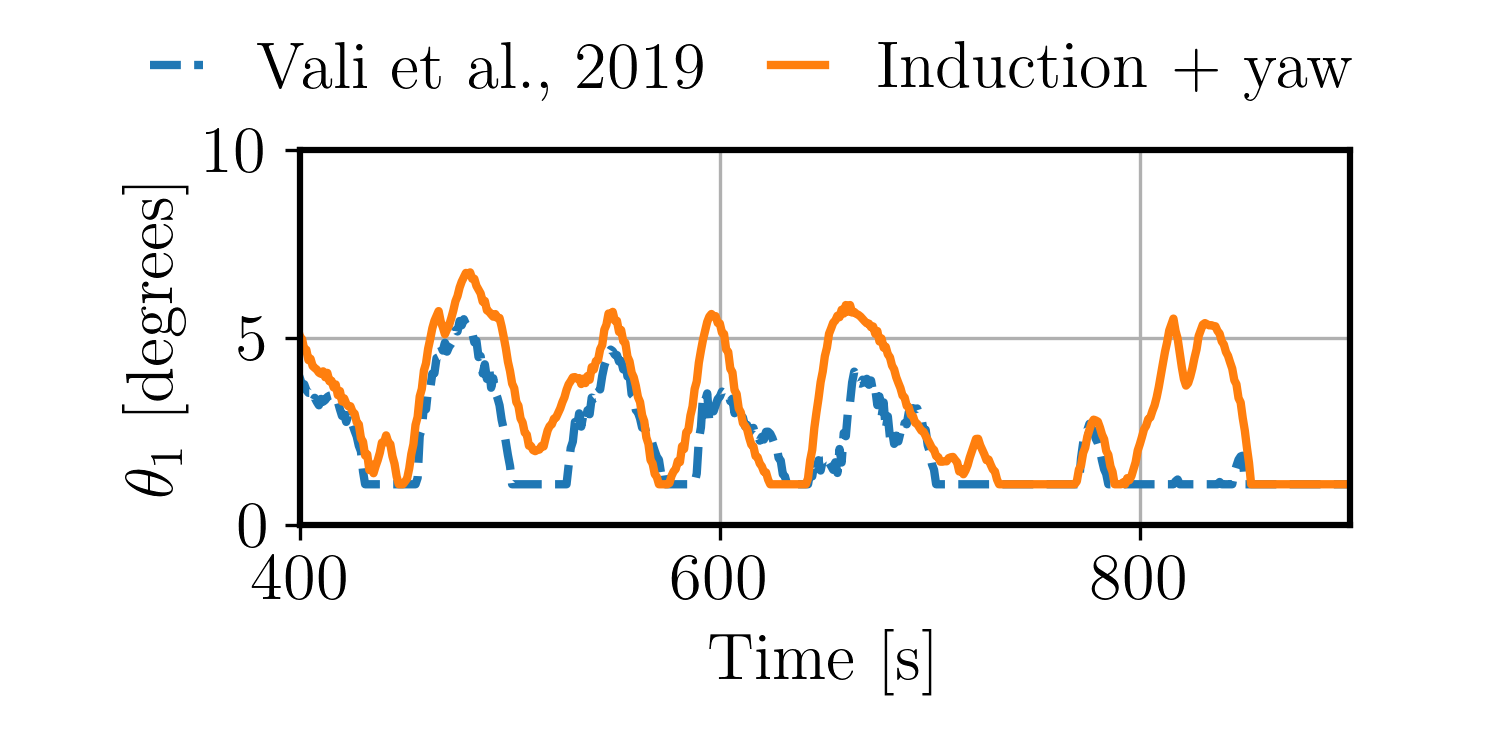}
            \caption{Wind turbine 2}
         \label{fig6b}
     \end{subfigure}

     \begin{subfigure}[b]{0.405\textwidth}
         \centering
            \includegraphics[width=\textwidth]{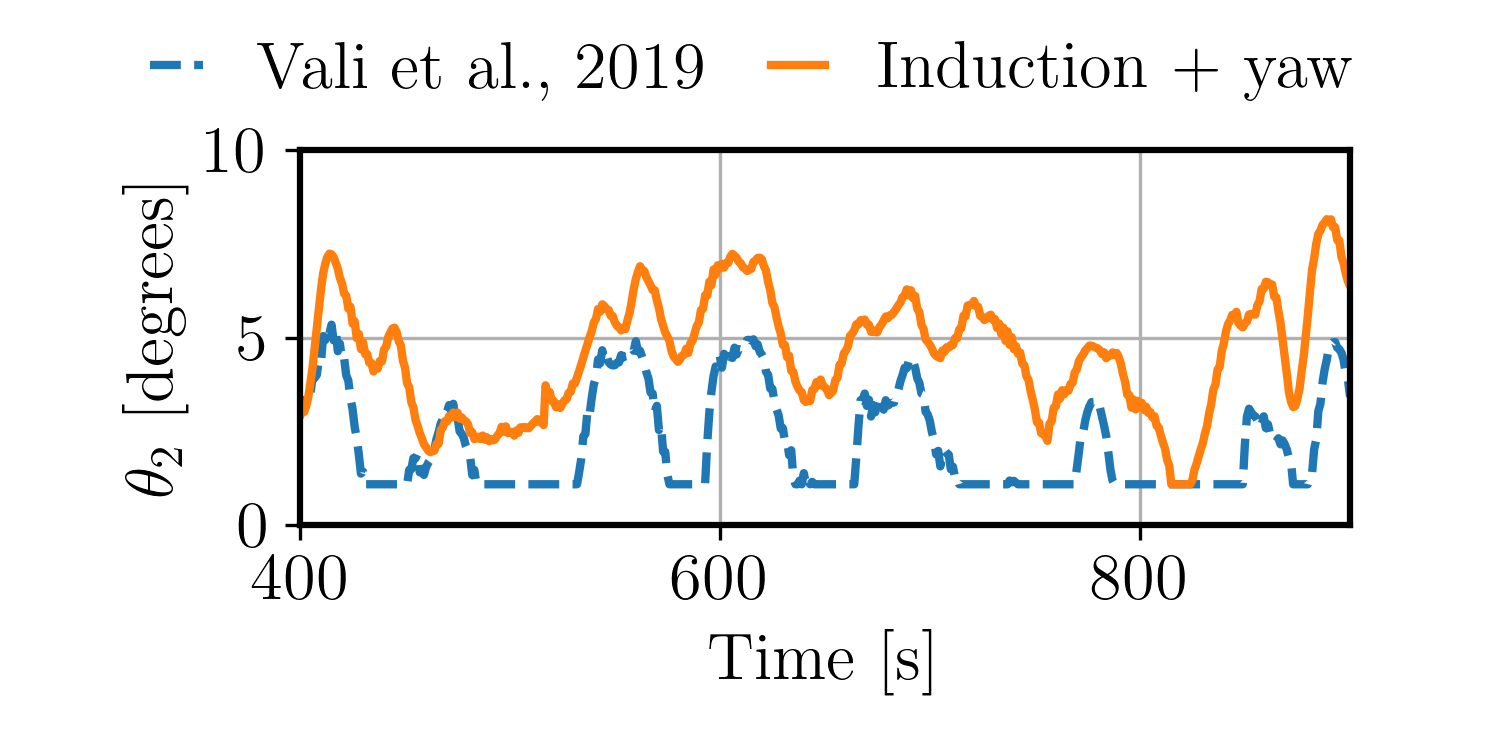}
            \caption{Wind turbine 3 (downstream)}
         \label{fig6c}
     \end{subfigure}
 \caption{Time series of pitch angles requested by the wind turbine controllers for the proposed (solid orange line) and benchmark (dashed blue line) controllers.}
 \label{pitch}
\end{figure}

In order to better understand how the local power margin is increased by the new method, the pitch angles commanded by the wind turbine controllers are plot in fig.~\ref{pitch}.

For a standard curtailment derating strategy, larger power reserves are obtained for larger absolute differences between the commanded pitch angle and the optimal value. Figures~\ref{fig6b} and~\ref{fig6c} show that waked turbines display the highest margin increase compared to the benchmark case, due to the lowered impact of the impinging wakes. On the other hand, the most upstream wind turbine (see fig.~\ref{6a}) generally displays a lower margin with the new control strategy, because of its yaw misalignment. Nevertheless, for the benchmark controller, the frequent saturation of the downstream turbines number 2 and 3 forces the upstream turbine number 1 to compensate, and in these conditions its margin drops relatively to the new proposed formulation.

\begin{figure}[htbp!]
\centering
\includegraphics[width=0.425\textwidth]{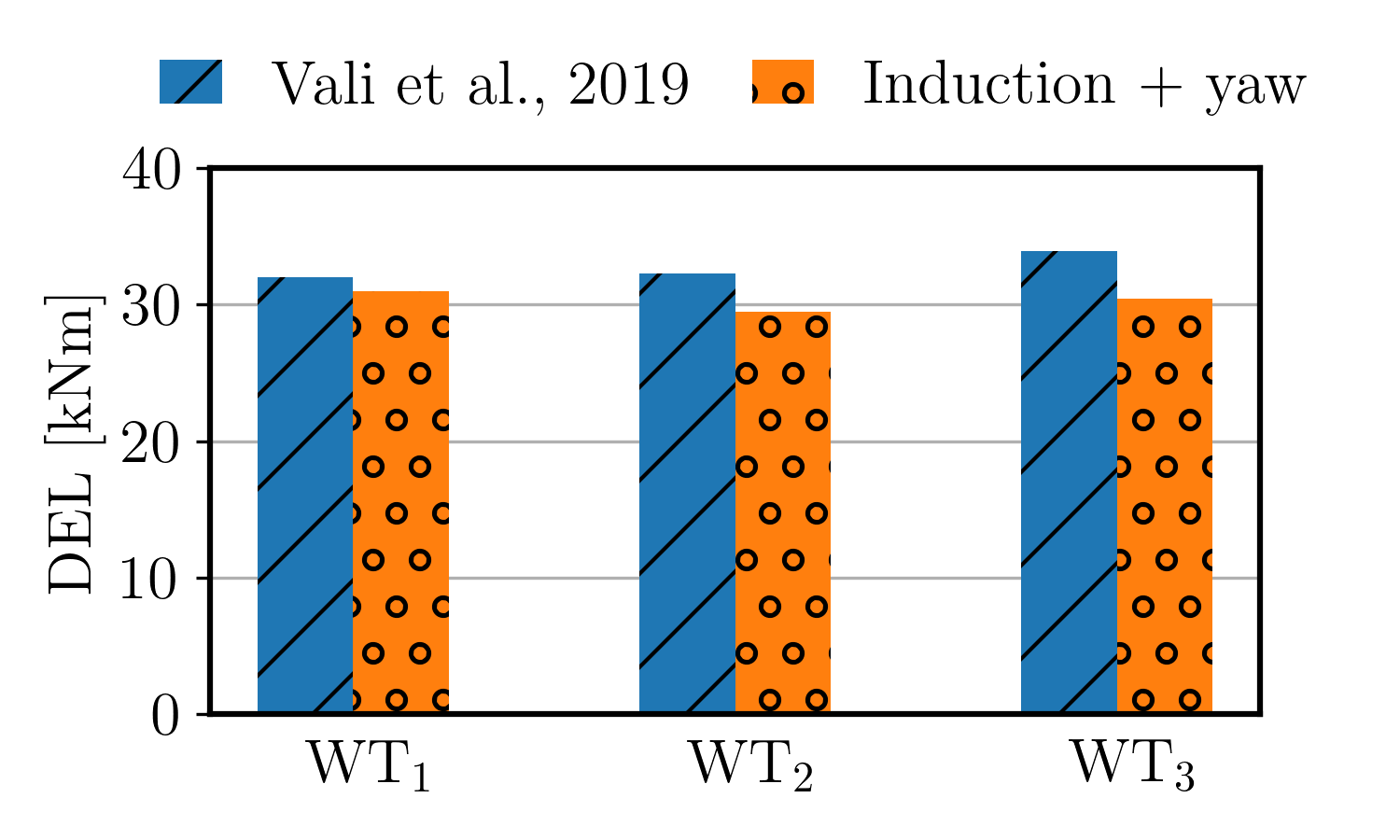}
\caption{Tower-base fore-aft bending moment DELs for  the proposed (orange bars) and benchmark (blue bars) controllers.}
\label{del}
\end{figure}

Finally, the effect of the new methodology on loads is briefly considered. Fig.~\ref{del} shows the damage equivalent loads (DEL), computed by rainflow counting (\cite{rainflow}), for the tower base fore-aft bending moment of each turbine.

Results indicate that the new control strategy reduces fatigue compared to the benchmark one. These results can be explained by the fact that the benchmark controller is unable to maintain load balancing within the farm in high-power-demand conditions, due to the frequent saturation events. Conversely, the new controller reduces the extent of the saturation phenomena, thereby suppressing the abrupt controller actions that are responsible for high-amplitude fatigue cycles.

\section{CONCLUSIONS}

A new wind farm control methodology for power tracking was presented. The methodology combines wake steering and induction control with the aim of maximizing the lowest power margin within a wind farm. The implementation is based on a slow-rate open-loop optimal set-point scheduler, combined with a fast feedback loop corrector. Compared to a  state-of-the-art benchmark, the new methodology is capable of reducing the root-mean-square of the tracking error  in conditions of power demand close to the maximum capacity of the plant. In such conditions, the fatigue of the individual wind turbines is also mitigated, because of less frequent saturation phenomena.

\addtolength{\textheight}{-2.5cm}

\section*{ACKNOWLEDGMENT}
The authors acknowledge the support of the German Federal Ministry for Economic Affairs and Climate Action (BMWK) through the \textit{PowerTracker} project. The authors express their appreciation to the Leibniz Supercomputing Centre (LRZ) for providing access and computing time on the SuperMUC Petascale System under Projekt-ID pr84be ``Large-eddy Simulation for Wind Farm Control''.

\bibliographystyle{IEEEtran}
\bibliography{IEEEabrv,root.bib}

\end{document}